\documentclass[twocolumn,groupedaddress,aps,prl,floatfix]{revtex4}

\usepackage{graphicx}  
\begin{document}

\title{A simple asynchronous replica-exchange implementation}
\author{Giovanni Bussi}
\email{gbussi@unimore.it}
\affiliation{Dipartimento di Fisica, Universit\`a di Modena e Reggio Emilia, \\
                 and CNR-INFM Center on nanoStructures and bioSystems at Surfaces (S3), \\
                      via Campi 213, 41100 Modena, Italy}
\date{\today}

\begin{abstract}
We discuss the possibility of implementing asynchronous replica-exchange
(or parallel tempering) molecular dynamics.
In our scheme, the exchange attempts are driven by asynchronous messages sent by one
of the computing nodes, so that different replicas are allowed to perform a different
number of time-steps between subsequent attempts.
The implementation is simple and based on the message-passing interface (MPI).
We illustrate the advantages of our scheme with respect to the standard
synchronous algorithm and we benchmark it
for a model Lennard-Jones liquid on an IBM-LS21 blade center cluster.
\end{abstract}

\maketitle

Parallel tempering is a popular method used to enhance sampling in Monte Carlo
or molecular dynamics simulations~\cite{swen-wang86prl,hans97cpl,sugi-okam99cpl},
with applications ranging from molecular biology to statistical physics.
In parallel tempering simulations, many replicas of the system are simulated
at different temperatures and their temperatures are exchanged in a Monte Carlo
fashion. The algorithm is very flexible, and can be extended to
simulation parameters other than the temperature
(see e.g.~Refs.~\cite{fuku+02jcp,jang+03prl,liu+05pnas}).
In this generalized form the method is usually referred to as \emph{replica exchange}.
Furthermore, parallel tempering (or replica exchange)
can be easily
combined with other enhanced-sampling methods~\cite{sugi+00jcp,buss+06jacs}.

A very important feature of replica-exchange simulations is that they
allow exploiting large parallel machines:
the standard implementation is done assigning a different
computing node to each replica, so that inter-process communications
are only needed when attempting an exchange move, which is typically
done with a prefixed stride.
Moreover, parallel tempering can be coded in such a way that temperatures are
exchanged instead of coordinates. In this case, the amount of data that are
transmitted between computing nodes 
is negligible, resulting in a theoretically perfect scalability.

However, practical problems arise when the number of replicas grows. In fact,
for hundreds of computing nodes, even the overhead of synchronization
can be large, especially when using short strides between subsequent attempts
as suggested e.g.~in Ref.~\cite{sind+08jcp}.
Furthermore, there is no warranty that the different nodes will take
the same time to perform the same number of steps.
As an example,
the frequency of updates of the neighbor list depends on the diffusion
coefficient~\cite{alle-tild87book}, which in turns depends
on the temperature, so that high temperature simulations are expected
to be slower.
The situation is even worse when heterogeneous machines are used.

A possible solution for these problems is the adoption of
serial methods~\cite{mari-pari92el,hage+07jpcb}.
In this Paper we will discuss a different approach, namely
the implementation of replica exchange in an
asynchronous manner.
The present implementation is simple and based on the MPI library~\cite{mpi}.
It is designed as a master-slave scheme,
where a node is asynchronously driving the exchanges
for the other replicas.
To the best of our knowledge, the only other attempt to implement asynchronous
parallel tempering is the
SALSA framework introduced in Refs.~\cite{zhan+06proceedings,gall+08jcc}.

\section{Implementation}
In our implementation temperatures are exchanged rather than coordinates.
This allows to minimize the amount of data that has to be communicated, at the price
of a slightly more complex analysis of the resulting trajectories.
Our asynchronous algorithm consists of the iteration of the following steps:
\begin{itemize}
\item Each node sends (asynchronously) its temperature to the master node,
      then starts to integrate molecular dynamics.
\item When $N_x$ steps are elapsed since the last exchange attempt, the master node
      collects the temperatures of all the other nodes, sorts them, and establishes the exchange pattern
      (i.e.~which pairs of nodes should attempt the exchange). Then, it
      sends (asynchronously) a message to each node (including itself)
      containing the rank of the partner node.
\item At every step, each node checks if a message arrived from the master. If the answer is negative,
      computation is continued.
      Otherwise, a message is sent (asynchronously) to the partner node, containing the actual temperature,
      potential energy and an uniformly distributed random number.
      Then, the node waits for the corresponding message from the parter node.
      Once both temperatures and energies are available, the random number generated by one of the two nodes
      (arbitrarily fixed to be that with the lowest rank)
      is used to establish if the exchange is successful using the standard Metropolis rule~\cite{sugi-okam99cpl}.
      In case of positive answer, the temperature and velocities
      are properly scaled.
\end{itemize}
As a further improvement, the rank of the master node is changed at each exchange, so that
the master overhead (sorting replicas and sending/receiving many messages) is distributed
over the full set of nodes.

Only two MPI calls need to be done in a blocking way: (a) the collection of temperatures
done by the master node, since all the temperatures should be known to prepare the exchange pattern;
(b) the reception of the triplet of temperature, energy and random number, which is needed
to determine whether the temperature has to be changed or not.
All the other calls to the MPI library are asynchronous and allows for overlap
of communication and computation.
The scheme can be trivially extended to support non-nearest neighbor exchanges.
When compared with the SALSA framework~\cite{zhan+06proceedings,gall+08jcc},
our scheme is simpler and has the advantage that it is based on a standard
MPI library and do not require multi-thread capabilities.

As a reference, we also implement a synchronous version, where the collection of the temperature
is done at every node with the prefixed stride using a MPI\_ALLGATHER call.
Here all the nodes are able to decide the exchange pattern and perform the
exchange after the same prefixed number of steps.

\section{Example}

We test our implementation using a Lennard-Jones liquid. This system
does not exhibit significant barriers, and parallel tempering is in principles
not needed to properly sample the phase space.
However, the diffusion of the replicas in temperature
space is similar to what is usually observed for simulation of solvated molecules,
so that it can be used as a meaningful test case.
We simulate 4000 particles at density $\rho=0.844$,
with cutoff in the interparticle interaction at 2.5.
Throughout this section we use Lennard-Jones units for distances, energies and time.
A neighbor-list containing pairs up to a distance 3.0 is used,
and it is updated when the displacement of at least one particle
is larger than half the skin width~\cite{alle-tild87book}.
The integration time-step is 0.005, and temperature is controlled using stochastic velocity rescaling~\cite{buss+07jcp}.
$N_r$ replicas are simulated at exponentially distributed temperatures in the
range from $T_{min}=0.7$ to $T_{max}=T_{min}\times1.0275^{N_r-1}$, which gives an exchange acceptance
of approximately 30\% irrespectively of $N_r$.
In the asynchronous implementation,
exchanges between replicas are attempted when the master node has done $N_x$ steps since the last attempt.
Since the number of steps performed by the other nodes is not prefixed, the actual average
number of steps between attempts $\langle N_x\rangle$ is not equal to $N_x$.
On the other hand, when the synchronous implementation is used the same number of steps is performed by
all the nodes between subsequent attempts, so that $\langle N_x \rangle=N_x$.
All the calculations are performed on the BCX cluster at CINECA.

\begin{figure}
\begin{center}
\includegraphics[width=0.45\textwidth,clip]{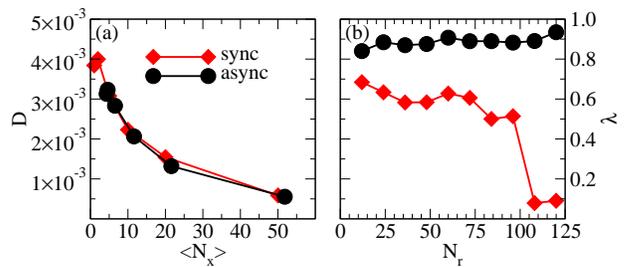}
\end{center}
\caption{
(a) Diffusion coefficient $D$ in the log-temperature space, as a function of the average number of waiting steps
$\langle N_x\rangle$.
(b) Parallel efficiency $\lambda$, as a function of the number of replicas $N_r$.
Both quantities are plotted for the synchronous (diamonds) and asynchronous (circles) implementations.
See text for details.
\label{fig1}
}
\end{figure}

\subsection{Diffusion in temperature space}
A very important parameter in replica-exchange simulations is the speed at which
each replica diffuses in the temperature space. This speed is inversely related
to the average time required for a replica to cross the full temperature range from
the lowest to the highest temperature.
We here calculate the diffusion coefficient of the logarithm of the temperature
in the case $N_r=48$ for different choices of $N_x$ in both the synchronous and asynchronous
implementation. The results are plotted in Fig.~\ref{fig1}(a) as a function of the observed
exchange stride $\langle N_x\rangle$. As it is seen, diffusion in both cases is equivalent and
only depends on the effective exchange stride. Moreover, as it has been observed
in Ref.~\cite{sind+08jcp}, the optimal performance is obtained by using an exchange stride
which is as small as possible. 

\subsection{Scalability}
We then consider the overhead due to the exchanges as the number of replicas $N_r$ grows.
To this aim we fix $N_x=5$ for the synchronous implementation and $N_x=1$ for the
asynchronous implementation. For the latter, this choice gives a practical 
waiting time of $\langle N_x\rangle\approx5$, so that the two algorithms are comparable.
We then define the parallel efficiency $\lambda$ as the ratio between the total number of
steps performed by all the replicas in a prefixed wall-clock time
and the same number as obtained
during a simulation without exchanges. In practice, this number should be equal to 1 if there was no
computational overhead, and decreases towards zero as the overhead increases.
In Figure~\ref{fig1}(b) we plot the efficiency $\lambda$ as a function of $N_r$.
It is clearly seen that the synchronous implementation is much less efficient than the asynchronous one,
and that it gets slower and slower as $N_r$ increases.
On the other hand, the efficiency of the asynchronous implementation is very close to 1 even when
more than a hundred replicas are used.

\section{Conclusions}

We have shown an asynchronous implementation of replica exchange molecular dynamics,
where the exchange attempts are driven by external commands sent by one of the replicas.
This implementation is based on the MPI library.
Its performance is excellent and allows for a significant saving of computational effort
in simulations with hundreds of replicas. The code is available on request.

\acknowledgments
The author is grateful to Carlo Cavazzoni for carefully reading the manuscript.
The CINECA supercomputing center is acknowledged for the computational time.

\end{document}